\documentclass[aps,prd,amsmath,amssymb,twocolumn]{revtex4}
\pdfoutput=1
\usepackage{graphicx}
\usepackage{color}
\newcommand{\beq}{\begin{eqnarray}}
\newcommand{\eeq}{\end{eqnarray}}

\newcommand{\bmp}{\noindent\begin{minipage}{16cm}}
\newcommand{\emp}{\end{minipage}\vskip 7mm} 

\usepackage{dcolumn}
\usepackage{bm}
\usepackage{bbm}
\usepackage{subfigure}
\usepackage{pxfonts}

\usepackage{epsfig}

\definecolor{rossoCP3}{cmyk}{0,.88,.77,.40}

\def\lsim{\mathrel{\rlap{\lower4pt\hbox{\hskip1pt$\sim$}}
    \raise1pt\hbox{$<$}}}                
\def\gsim{\mathrel{\rlap{\lower4pt\hbox{\hskip1pt$\sim$}}
    \raise1pt\hbox{$>$}}}                

\usepackage{fancyhdr}
\begin{document}
\title{\Large  \color{rossoCP3} Nonperturbative Results for Yang-Mills Theories}
\author{Francesco {\sc Sannino}$^{\color{rossoCP3}{\varheartsuit}}$}
\email{sannino@cp3.sdu.dk} 
\affiliation{
$^{\color{rossoCP3}{\varheartsuit}}${ CP}$^{ \bf 3}${-Origins}, 
Campusvej 55, DK-5230 Odense M, Denmark. }
\author{Joseph {\sc Schechter}$^{\color{rossoCP3}{\clubsuit}}$}
\email{schechte@phy.syr.edu} 
\affiliation{\mbox{
$^{\color{rossoCP3}{\clubsuit}}$
Dept. of Physics, Syracuse University, Syracuse, New York 13244-1130, USA}}
 
\begin{abstract}
      Some non perturbative aspects of the pure SU(3) Yang-Mills theory are 
      investigated assuming a specific form of the beta function, based on
      a recent modification by Ryttov and Sannino of
      the known one for supersymmetric gauge theories. The characteristic 
      feature is a pole at a particular value of the coupling constant, $g$.
      First it is noted, using dimensional analysis, that physical quantities
      behave smoothly as one travels from one side of the pole to the other.
      Then it is argued that the form of the integrated beta function $g(\mu)$,
      where $\mu$ is the mass scale, determines the mass gap of the theory.
      Assuming the usual QCD value one finds it to be 1.67 GeV, which is in surprisingly good
      agreement with a quenched lattice calculation. A similar calculation is made 
      for the supersymmetric Yang-Mills theory where the corresponding beta function is
      considered to be exact. 
       \\[.1cm]
      {\footnotesize  \it Preprint: CP$^3$-Origins-2010-36}
\end{abstract}

\maketitle
 \thispagestyle{fancy}
\section {Introduction}

   The exact beta function of a gauge theory is generally considered to
   contain many non-perturbative {\it secrets} of the gauge theory behavior. Unfortunately 
   it seems to be only computable analytically in perturbation theory. Physically it is related 
   to the trace anomaly, or the non-zero value of the divergence of the scale (dilation)
  current. We are specializing to massless theories here so it represents a violation of the classical
  result.
  
    Another object in  gauge theories with massless fermions, the divergence of the U(1) axial 
    vector current, should be zero in the classical limit but is known not to vanish at the quantum
    level (axial anomaly). Most interestingly it has the property that there are no corrections \cite{ab}
    beyond the one loop level so it can be considered to be known exactly.

Now, in supersymmetric gauge theories, it was found \cite{fz} that both the trace anomaly 
and the axial anomaly (actually together with the divergence of a special 
superconformal current) belong to the same chiral multiplet and hence should
be somehow related to each other. There was a lot of discussion about the meaning of
this feature - for example is there a contradiction between an exact one loop result for the 
axial anomaly and a result containing all orders of coupling constant for the
trace anomaly - and finally it was realized that there could be compatibility, with 
the exact beta function being determined from  the axial anomaly. This was shown 
both for the supersymmetric Yang-Mills theory \cite{jl} and for the supersymmetric theory containing
also fermions not belonging
to the adjoint representation \cite{nsvz}. A characteristic feature of these exact beta
functions is a pole at a particular value of coupling constant, {g} in the beta 
function $\beta(g)$.

    Recently, Ryttov and Sannino \cite{Ryttov:2007cx} conjectured such {\it all order} beta  
functions for ordinary $SU(N)$ (non supersymmetric) gauge theories both with and without 
fermions belonging to arbitrary representations based on analogy to the 
supersymmetric case. The generalization, by one of the authors, for the $SO$, and $Sp$ gauge groups appeared in \cite{Sannino:2009aw} and for chiral gauge theories in \cite{Sannino:2009za}. These 
beta functions were found to satisfy known consistency conditions at second order and 
to work well in many interesting applications to working technicolor models reviewed in\cite{Sannino:2009za}. These beta functions also feature a pole at a particular value of g.

     In the present note, we study some non-pertubative consequences of the conjectured
     beta function for the simple SU(N) Yang-Mills theory; already \cite{Ryttov:2007cx} the conjectured
     beta function had been 
      seen to give a reasonable picture in the asymptotically free perturbative region. In section II 
      we use dimensional analysis to investigate the running with scale of physical quantities with 
      various engineering dimensions. This involves $g(\mu)$ but it was noted that physical quantities
      change smoothly as g goes through the pole value. 
      
      In section III, we investigate the integration of the defining equation for the
      beta function which yields an explicit expression for $g(\mu)$. We consider all Yang-Mills 
      theories in the sense that a full range of values for the coupling constant at a 
      reference mass are considered. An amusing feature is seen to arise: the solutions for $g(\mu)$
      do not allow $\mu$ to be lower than a certain value. We interpret this as the measure
      of the mass gap for the Yang-Mills theory. For the usual value of the QCD coupling constant,
      with N=3, our predicted value of 1.67 GeV is seen to be in good agreement with a valence
      lattice calculation. We also compute the mass gap for the supersymmetric SU(3) and SU(N)
      Yang-Mills theories. In the latter case, the starting beta function does not require any conjecture.
      
      Section IV contains a brief summary and discussion.

\section{General Scaling Results for Yang - Mills Theories}
Consider a renormalization group invariant quantity $H$ of mass dimension $D$ in any
theory with a single field and a single coupling constant $g$.
Define $\mu$ to be the renormalization scale and $g(\mu)$ 
to be the value assumed by the coupling constant at that scale. Since $H$ must
be a measurable physical quantity, dimensional analysis implies that 
it has the form, 
\begin{equation}
H = \mu^D \, {\cal H}\left[ g(\mu) \right] \ ,
\label{diman} 
\end{equation}
where ${\cal H}\left[g(\mu)\right]$ is some function of the coupling constant.
Of course, for any given theory there are many different interesting 
quantities $H$. One may, as usual, introduce a characteristic {\it invariant scale}, 
$\Lambda$ for the theory by defining a {\it particular} $H$ to be $\Lambda^D$.
 By differentiating both sides of Eq.(\ref{diman}) with respect to 
  $\log(\mu)$, the natural log, and recognizing that the left hand side is independent of $\mu$, 
   we obtain the main equation: 
\begin{equation}
D\, {\cal H} + \frac{\partial {\cal H }}{\partial g} \beta(g) = 0\
 , \quad {\rm with} \quad \beta(g) = \frac{\partial g}{ \partial \log(\mu)} \ .
 \label{main}
\end{equation}
By integrating this equation one immediately finds: 
\begin{equation}
\log \left( \frac{{\cal H}}{ \cal H}_0\right)  = - D \int_{g_0}^g\frac{dg}{\beta(g)} \ .
\label{log}
\end{equation}
Hence if the exact beta function were known as a function of $g$, one
could express any renormalization group invariant quantity $H$ as a 
known function of $g$ up to an arbitrary overall constant. 

Now, as discussed in the preceding section, a conjectured all orders beta function
for the ordinary Yang-Mills theory based on the known supersymmetric Yang-Mills 
theory all orders beta function was recently introduced in \cite{Ryttov:2007cx} by Ryttov and Sannino. It
was found to be consistent with other non perturbative approaches to the 
ordinary (non supersymmetric) gauge theories and hence to be a reasonable
model for further investigation.

This RS all orders beta function ansatz for the SU(N) Yang-Mills theory reads:
\begin{equation}
\beta_{YM} = - g^3 \frac{a}{1-b g^2}\  , \quad {\rm with}
 \quad a = \frac{11}{3} \frac{N}{(4\pi)^2}    \quad {\rm and} \quad  b = \frac{17}{11} \frac{N}{8\pi^2} \ . 
\label{RS}
\end{equation}
Using Eq.(\ref{RS}) in Eq.(\ref{log}) then yields:
\begin{equation}
H   = const \frac{\mu^D}{g^{\frac{Db}{a}}} \, \exp \left[-\frac{D}{2a\,g^2}\right] \ ,
\label{exp}
\end{equation}
where the overall constant is defined by
\begin{equation}
const = {\cal H}_0
 {g_0}^\frac{Db}{a}\exp \left[\frac{D}{2a\,g_0^2}\right].
\label{const}
\end{equation}
Here the subscript zero denotes the value corresponding to the lower 
limit of integration in Eq. (\ref{log}).
Of course, there is a different numerical constant for each choice of H. These 
numerical constants might be approximated by using perturbation theory at a
 large value of $\mu$, for example.
 
 For definiteness we list some possible interesting choices for $H$. 
 
 \begin{itemize}
 \item[ i.]{ The gluon condensate or vacuum expectation value of the trace of the energy 
momentum tensor. This has the {\it engineering} dimension $D=4$.}
 
\item[ ii.]{The glueball squared masses ($D$ = 2). We presume that there is
 a spectrum of glueballs with different spin-parities and masses.} 

\item[ iii.]{ The glueball-glueball scattering cross sections ($D$ = - 2).}
 
\item[ iv.]{ The coefficients, $a_n$ in the expansion of the partial 
 wave amplitudes for glueball scattering, $\Sigma a_n s^n$ in a 
 region of analyticity (Here D= -2n).}
 \end{itemize}
 
\begin{figure}[t]
\begin{center}
\subfigure[~Case $D=4$ ]{\includegraphics[width=7.5cm]{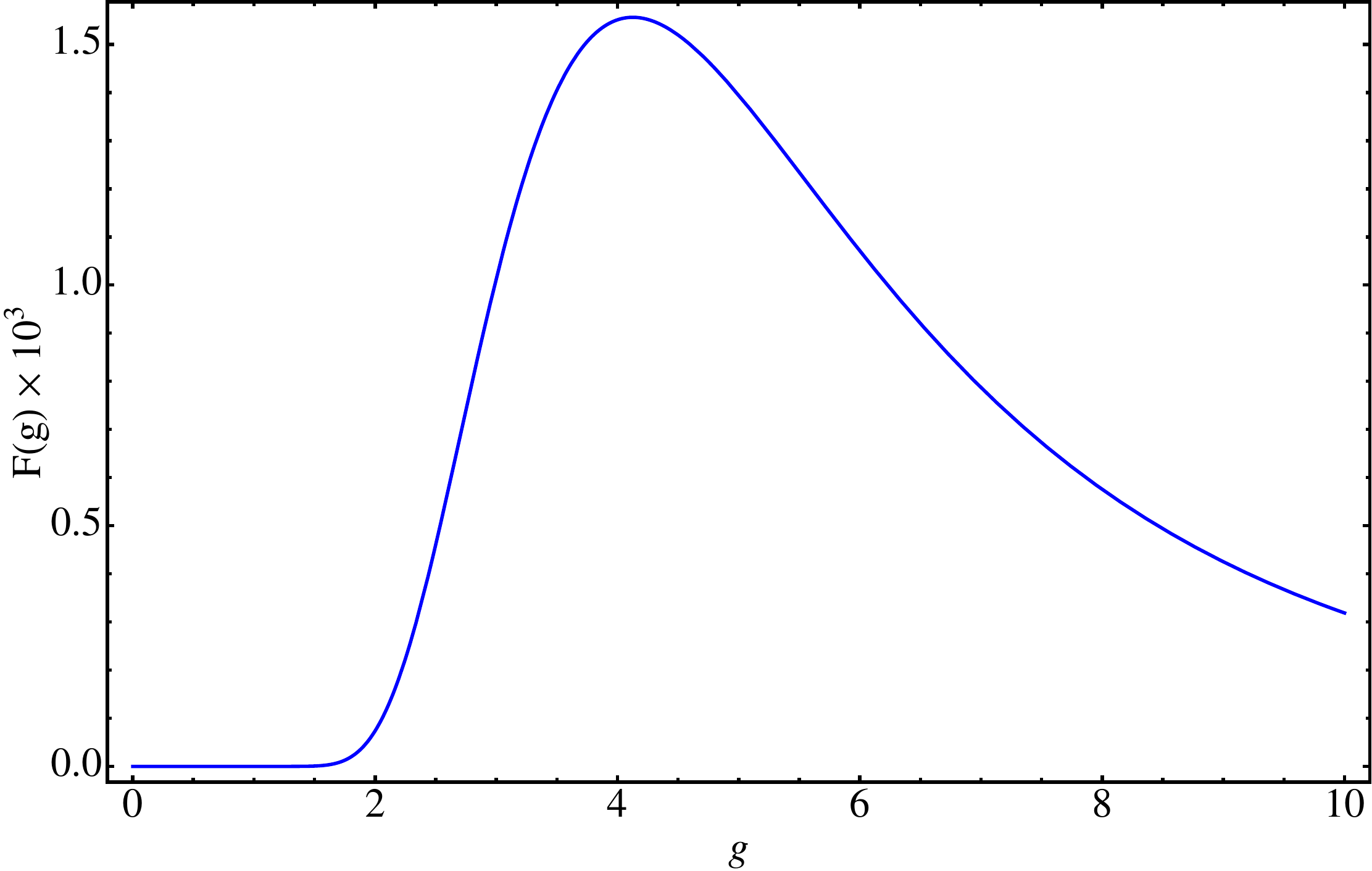}} \hfill
\subfigure[~Case $D=-2$]{\includegraphics[width=7.5cm] {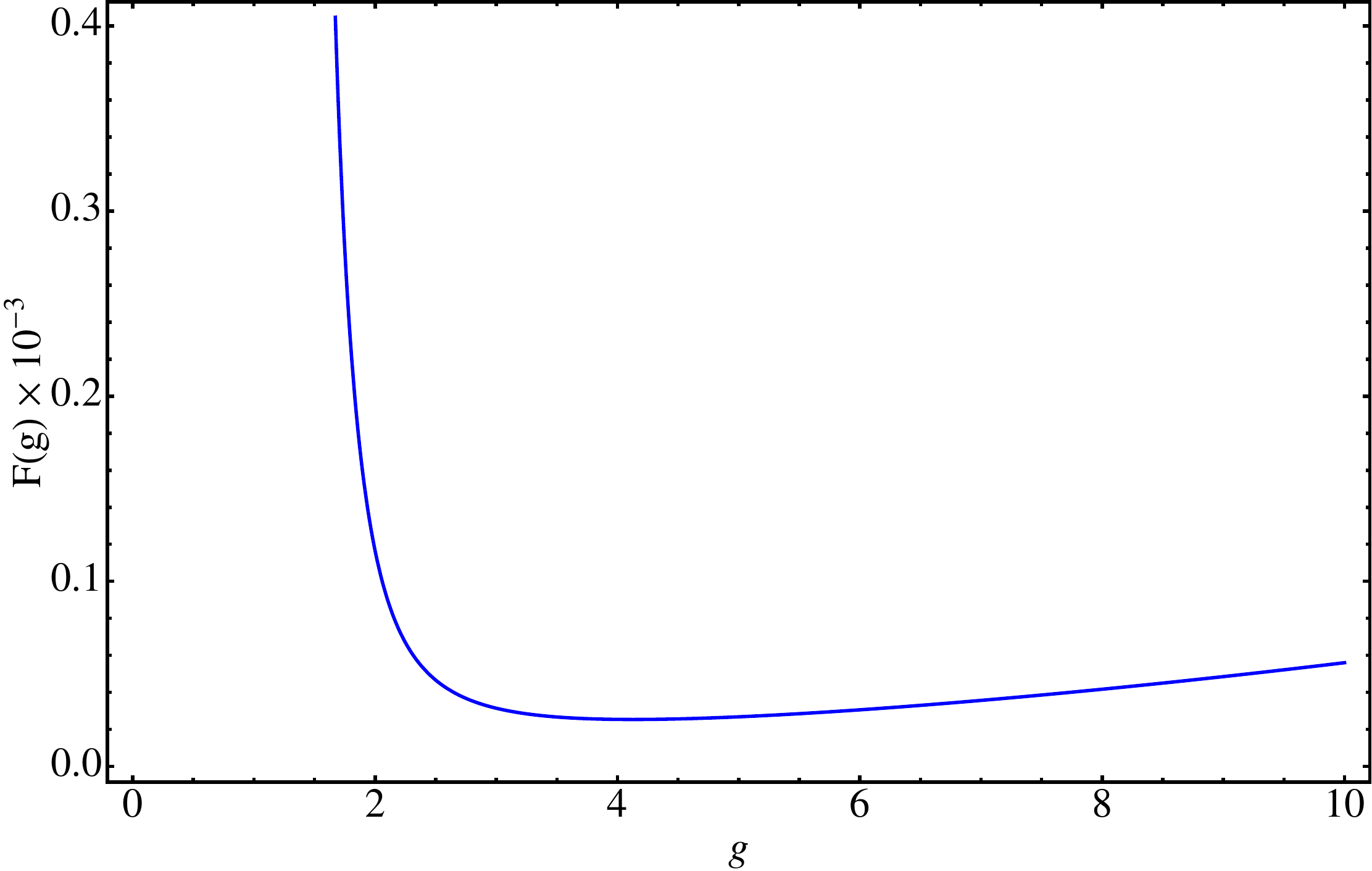}} \\
\caption{$F(g)$ as function of $g$ for $N=3$.} 
\label{FIG12}
\end{center}
\end{figure}

    To get an idea of the dependences of various physical quantities on
    the coupling constant $g$ we plot the characteristic factor,
        \begin{equation}    
    F(g) =  \frac{1}{g^{\frac{Db}{a}}} \, \exp \left[-\frac{D}{2a\,g^2}\right] \ , 
      \label{F}
      \end{equation}
 in Eq.(\ref{exp}) for the case $D=4$ in Fig.~\ref{FIG12} (a) one and for the case 
 $D=-2$ in Fig.~\ref{FIG12} (b) . For definiteness the choice $N$  equal to 3 is made.
 It is seen that each of these curves displays an 
 extremum at the same value of $g$. 
 
 Differentiating shows that this
 value is,
    \begin{equation}
    g^2 =1/b = \frac{88\pi^2}{51} \approx (4.127)^2.
    \label{extremum}
    \end{equation} 
Furthermore, the second derivative of $F(g)$ at this point is simply 
$-2DF(g)/ag^4$; this means that dynamical quantities with positive dimension $D$
will have a maximum at this point while those with negative D will have a
minimum at the same point. We also note the interesting fact that for a given value 
of the {\it physical} quantity $F$, there are two different values of $g$. The 
situation will be further explored by looking at the beta function. 

\section{Beta function and mass gap}

    A graph of the $N$ =3 Yang-Mills conjectured beta function in Eq.(\ref{RS})
  is shown in Fig.~2(a). From its shape one concludes that the origin, 
  $g$ =0 is an ultraviolet stable fixed point and that there is clearly a pole at $g = b^{-1/2}$. However, we observe from 
the Figs.~\ref{FIG12} (a) and (b) that the physical quantities, $H$ remain smooth as one goes 
  from one side of the pole to the other. 
 
 \begin{figure}[h]
\subfigure[~$\beta_{YM}(g)$ as function of $g$ for $N=3$. ]{\includegraphics[width=7.5cm]{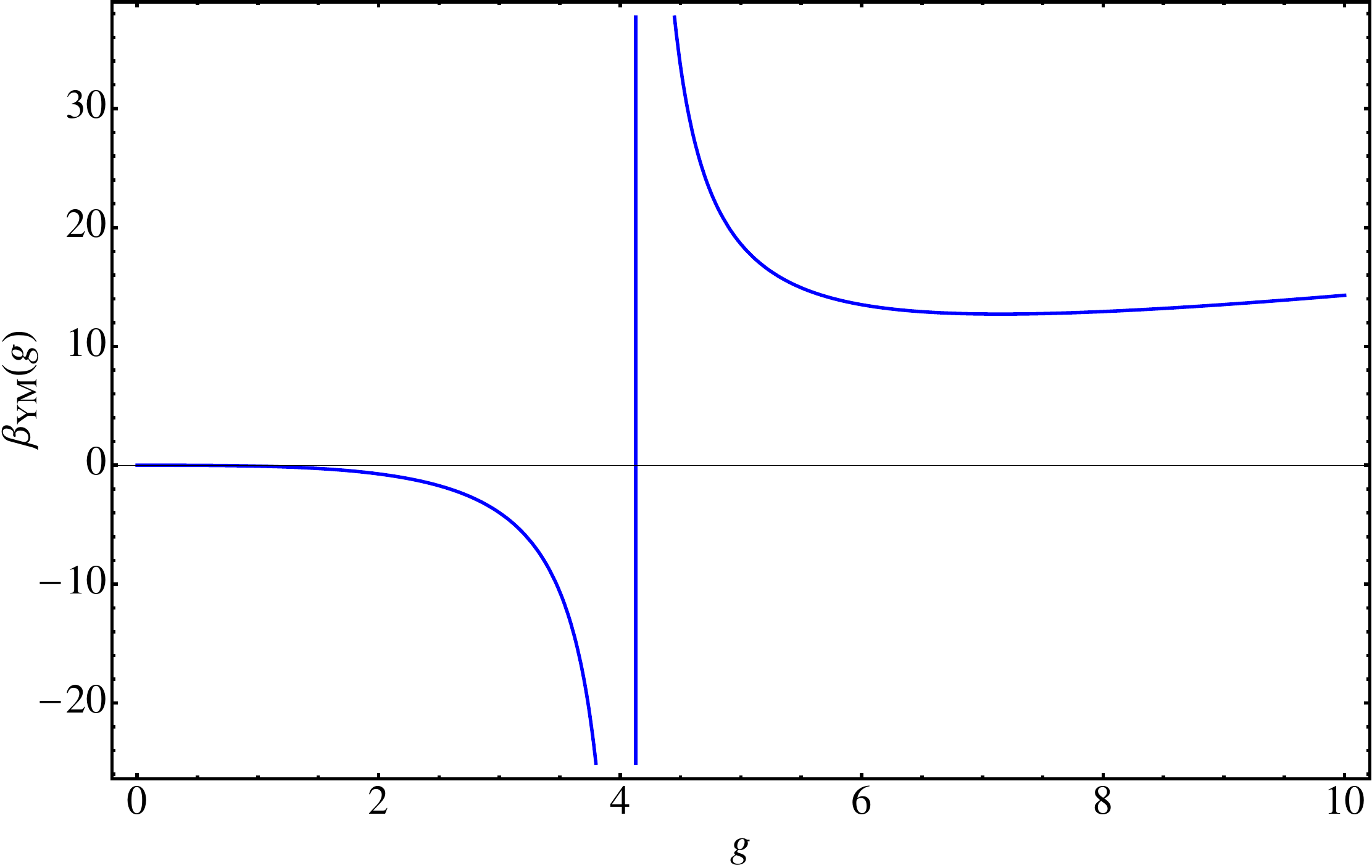}} \hfill
\subfigure[~ $\log\left(\mu/\mu_0 \right)$  as function of the coupling $g$ ]{\includegraphics[width=7.5cm] {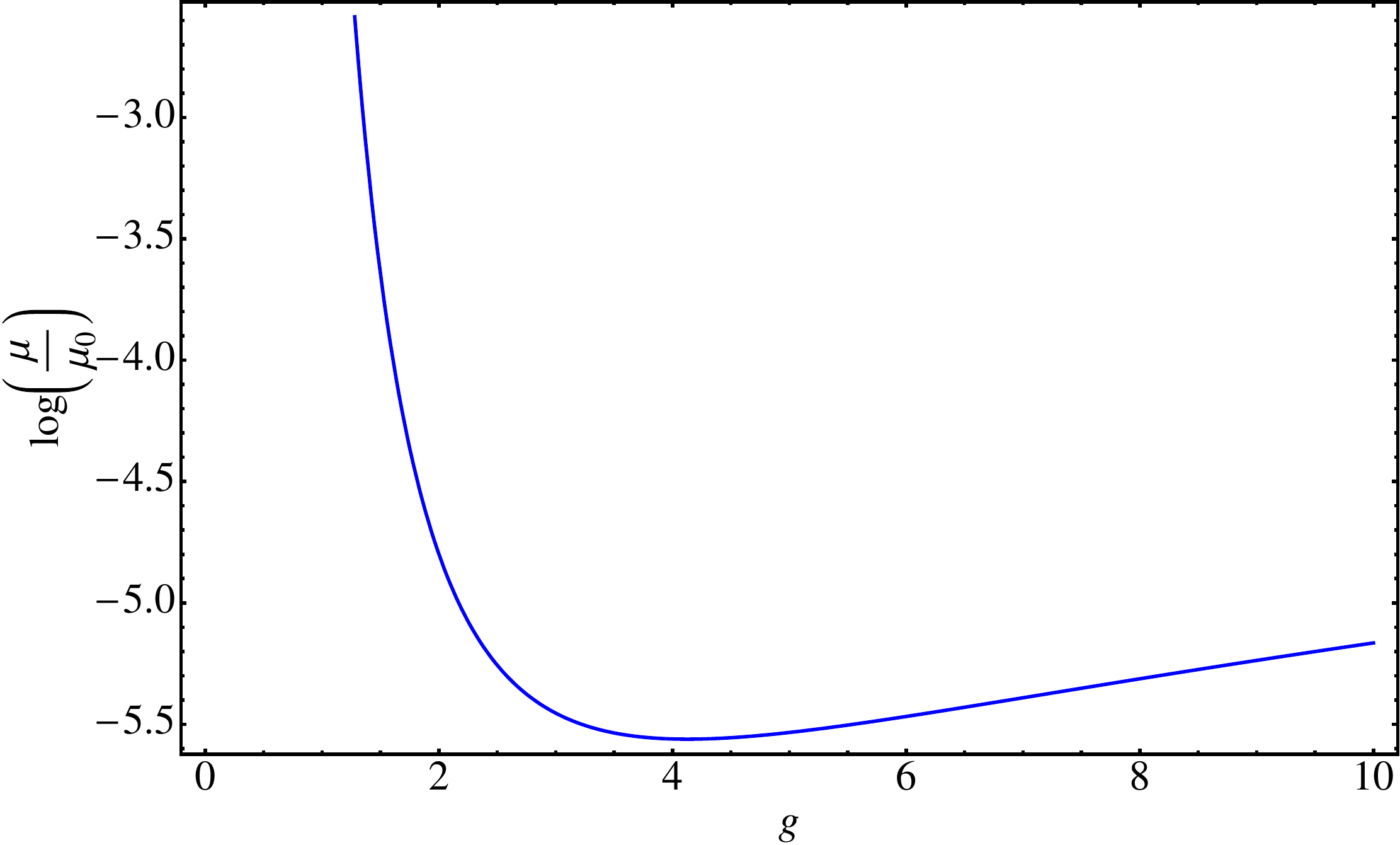}} \\
\caption{Plots of the YM beta function, panel (a), and of the solution for the running of the scale (b) as function of the coupling constant $g$ with $g_0=1$.} 
\label{FIGbeta}
\end{figure}%
%
%
      In order to discuss the coupling constant in a gauge theory, one chooses a 
      reference value $\mu_0$ of the energy scale and specifies (from experiment) the 
      value of the running coupling constant, $g(\mu_0)$ at this point. In the case of
      QCD, it is convenient to choose $\mu_0$ = $m_Z \approx$ 91.19 GeV. For illustration 
      we will also choose this value of $\mu_0$ as our reference scale. The value of the
      QCD coupling constant at this scale is measured to be about 1.228. In the present
      pure Yang-Mills case we are dealing with a hypothetical model so we are free to choose
      any value for $g(\mu_0)$. In fact it is better to allow  $g(\mu_0)$ to range
       over all possible values, which corresponds to describing all pure Yang-Mills 
       N=3 theories.
 
     Clearly, the first step is finding out how the coupling constant runs with the scale $\mu$.
     Integrating the second of Eq.(\ref{main}) yields the following
     relation between $g$ = $g(\mu)$ and $log(\mu)$ corresponding to any choice of
     $g_0$ = $g(\mu_0)$:
     
     \begin{equation}
 \log\left(\frac{\mu}{\mu_0} \right)=    \frac{1}{2a} \left( \frac{1}{g^2} - \frac{1}{g_0^2} \right)
  + \frac{b}{a}\log\left(\frac{g}{g_0} \right) \ . 
 \label{gruns}
 \end{equation}
 
     It is convenient to plot in Fig.~2(b)  $\log\left(\mu/\mu_0 \right)$ as a function of $g$ with the
      choice $g_0=1$. In the part of this plot to the 
     left of $g(\mu)=b^{-0.5} \approx$ 4.127, 
      $g(\mu)$ decreases as $\log(\mu)$ increases. That is the expected 
      asymptotically free behavior. On the other hand, to the right
      of this point, $g(\mu)$
       smoothly starts rising with increasing $\log(\mu)$. There is no
      discontinuity at the pole. Clearly, this behavior is the same as that shown
      for {\it physical quantities} in panels (a) and (b) of Figs.~\ref{FIG12}. The existence of a 
      smoothly connected different {\it phase} for the theory is intriguing.
      For the present, however, we will concentrate on the asymptotically free
      region.

      The different $N$ = 3 pure Yang-Mills theory which is defined by 
     $g_0=3$ yields the running coupling constant plotted in Fig.~3(a).
     The overall picture is substantially the same. In particular, the 
     asymptotically free region still corresponds to $g(\mu)<b^{-0.5}$.
     Similarly the theory characterized by $g_0=0.5$ is seen from the plot 
     in Fig.~3(b) to still have the same range in $g(\mu)$ for the asymptotically 
     free region, but with rather different $\mu$ values.
     
         Figures 2(b), 3(a) and 3(b) each illustrate that there is a particular
          (but different in each case) 
         value of $\log(\mu)$ below which the curve does not extend.
         This is peculiar since it would imply that the running coupling constant
         could not be {\it measured experimentally} in that region. The only way in which 
         this might be consistent is if the disallowed region would be lower than $M$,
         twice the mass of the lightest glueball state in the theory. Then it would be 
         below threshold and not accessible to experiment. That seems like a plausible 
         determination of the {\it mass gap} of the theory. It is clear that the value of such a mass gap 
         is determined by the ordinate of the point where $\log(\mu)$
         is minimum, i.e. where $g=b^{-0.5}$. Specifically,

 \begin{figure}[t]
\subfigure[~   $g_0=3$  ]{\includegraphics[width=7.5cm]{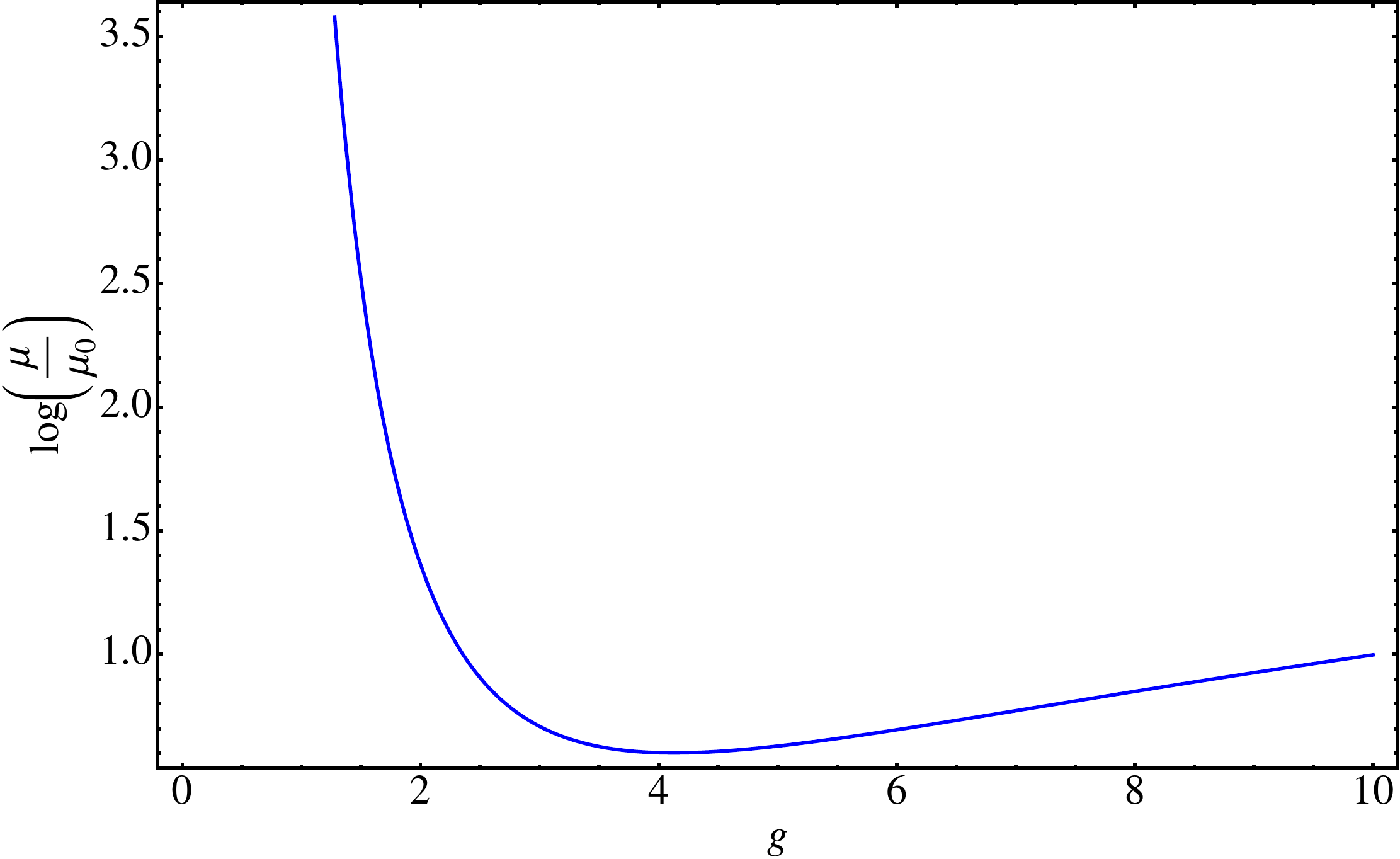}} \hfill
\subfigure[~ $g_0=0.5$ ]{\includegraphics[width=7.5cm] {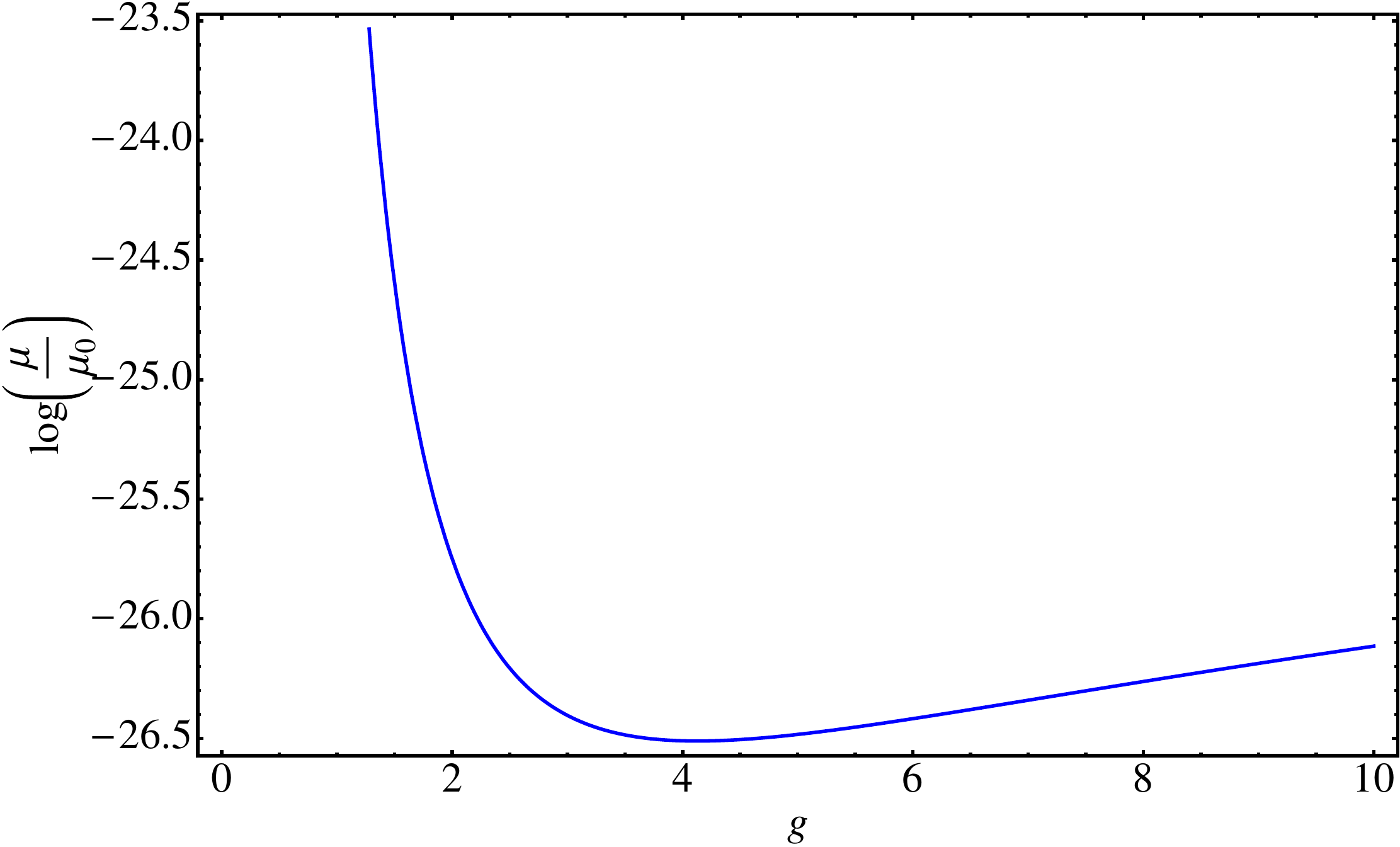}} \\
\caption{Solution for running scale vs. coupling constant $g$.  } 
\label{FIGbeta}
\end{figure}%

         \begin{equation}
         \log(M)=\log(\mu_0)+ \frac{1}{2a}\left( b -\frac{1}{g_0^2} \right)- \frac{b}{a}\log(g_0b^{0.5}) ,
         \label{massgap}
         \end{equation}
         
         where $g_0$ is the assumed value of the coupling constant at $\mu_0$ = 91.19 GeV. 
         
\begin{figure}[b]
\subfigure[~  $M$ (twice the mass gap) dependence on  $1 \leq g_0 \leq 10$  ]{\includegraphics[width=7.5cm]{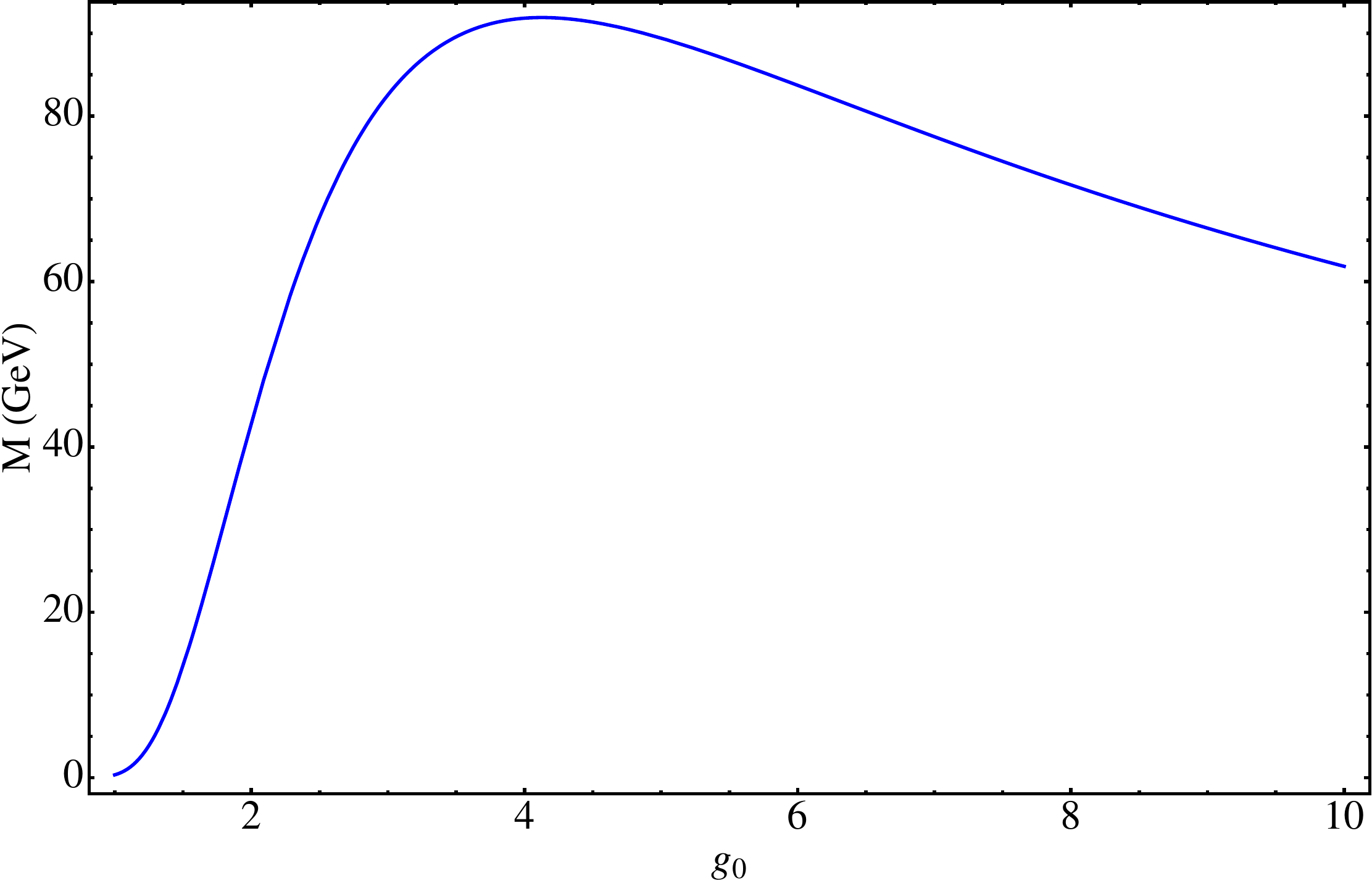}} \hfill
\subfigure[~$M$ (twice the mass gap) dependence on  $g_0 \leq 1$ ]{\includegraphics[width=7.5cm] {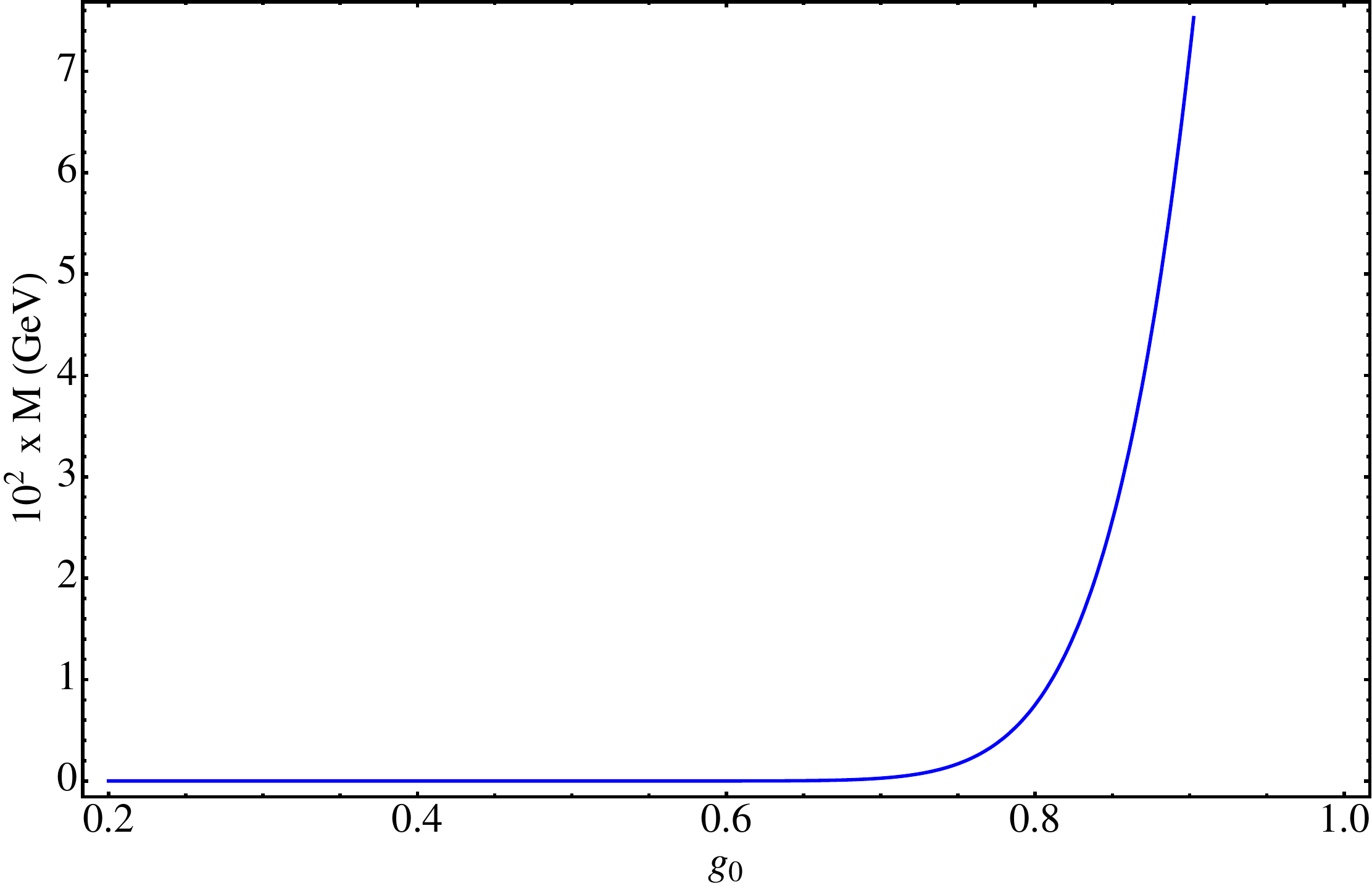}} \\
\caption{$M$ (twice the mass gap) dependence on  $g_0$ for $\mu_0 = 91.19$~GeV and $g = b^{-0.5}$.   } 
\label{FIGMassgap}
\end{figure}

         Note that the so defined mass gap depends on the particular N=3 Yang-Mills
          theory we are considering via its dependence on $g_0$. In Figure~\ref{FIGMassgap}~(a)  
          twice the mass gap, M in GeV is plotted for $g_0$
          ranging from 1 to 10. For orientation, the mass gap for $g_0$ slightly larger than
          one is seen to be in the GeV range.    The mass gap reaches a maximum when $g_0$ takes the {\it pole}  value, $1/b^{0.5}$ and declines somewhat for larger values of $g_0$.
           It seems reasonable that within the asymptotically free regime, the 
          mass gap increases with the strength of the coupling constant, $g_0$.
           For $g_0 <$ 1, the mass gap starts to decline extremely rapidly, as may be seen
           from Fig.~\ref{FIGMassgap}~(b).

\begin{figure}[h]
{\includegraphics[width=7.5cm]{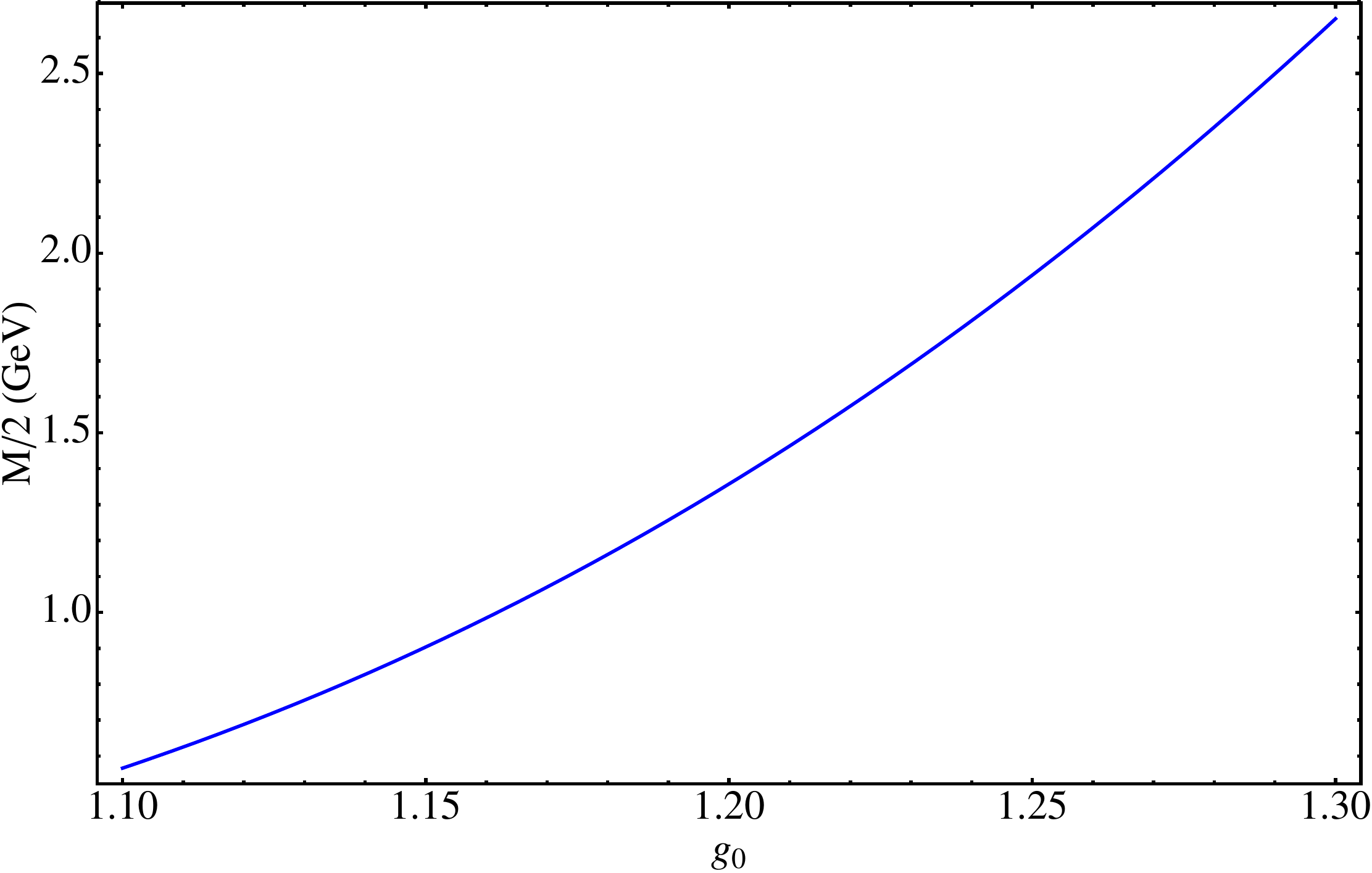}} 
\caption{Mass gap dependence on  $1.1 \leq  g_0 \leq 1.3$  for $\mu_0 = 91.19$~GeV and $g = b^{-0.5}$.   } 
\label{FIGgap}
\end{figure}

          To further test the reasonableness of our interpretation, we may try to predict
          the mass of the lightest glueball in this model. We would like the result 
          to be similar to the usually expected value (about 1.5 GeV) when we adopt
          the QCD value for the coupling constant, $g_0$ = 1.228, mentioned above.
          This would correspond to neglecting the effects of quark fields on the glueball
          mass. Since it is seen that there is a rapid dependence of $M$ on $g_0$ we give 
          in Fig.~\ref{FIGgap} 
          a {\it blow up} of the prediction for $M/2$ in the region of $g_0$ near the experimental one.
          This yields for the lightest glueball mass,
          \begin{equation}
          M/2 \approx 1.67~ {\rm GeV},
          \label{lightestgb}
          \end{equation}
          which does seem reasonable for an a priori prediction.
                    We may compare this value with the result of a lattice QCD 
          calculation employing a  {\it valence} assumption \cite{vw}. That calculation
          gave the result for the $0^+$ glueball which is found to be lightest:
          \begin{equation}
          m(0^+)= 1648 \pm 58 ~ {\rm MeV}.
          \label{latticemass}
          \end{equation}
          This embarassingly accurate agreement gives us at least some confidence
          in the correctness of the interpretation of the mass gap and the
          validity of the RS conjectured beta function. Other lattice results are also in agreement with the value quoted above \cite{Morningstar:1997ff,Lucini:2010nv}.                
          To sum up, the pole in the beta function does not produce any 
          singularity in the theory but seems to be the feature
          which generates the mass gap. It would be interesting to investigate
          the dependence of this result on the choice of renormalization scheme.

          If the conjectured pure Yang-Mills beta function could be shown to be 
          the same as the exact one, it would amount to showing that the theory has 
          the kind of mass gap defined above. 
     
              Actually, it is easy to see that this mass gap mechanism does not, in 
              general, require the RS conjecture if one goes beyond the ordinary 
              Yang-Mills model. That is because the supersymmetric Yang-Mills theory, which is 
               of course the starting point of the RS conjecture, is
              known to possess an exact beta function of the same form as 
              Eq.(\ref{RS}) but with somewhat different values of $a$ and $b$, namely: 
     
           \begin{equation}     
            a^{\prime} =  \frac{3N}{(4\pi)^2} \ , \quad
            b^{\prime} =  \frac{N}{8\pi^2}.
           \label{susyparam}
           \end{equation} 
 
\begin{figure}[t]
{\includegraphics[width=7.5cm]{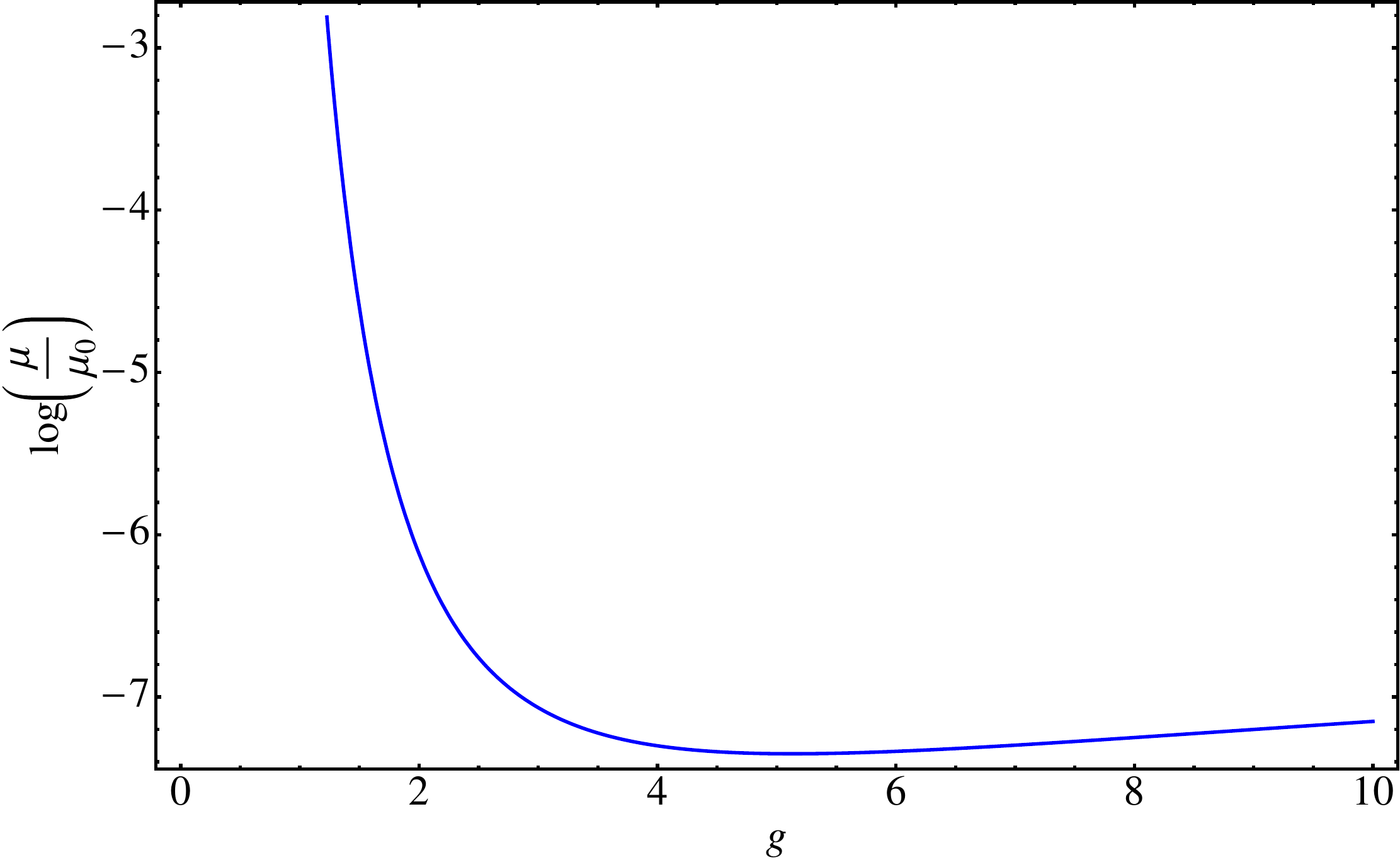}} 
\caption{Solution of the running of the scale as function of the coupling constant $g$ for the Super Yang-Mills theory.    } 
\label{FIGSUSY}
\end{figure}

              The numerical results are qualitatively similar if one assumes
          a similar value of the coupling constant. For example if we choose to
          specify the super Yang-Mills theory by taking a similar coupling constant 
          at the scale, $\mu_0 =m_Z$, we would get for the choice $g_0=1.0$, the curve
          shown in Fig.~\ref{FIGSUSY}, which describes the running of $g(\mu)$ with respect 
          to $\log(\mu)$ and can be compared with the curve in Fig.~2(b).
          
              We can compare the mass gap, $M^{\prime}/2$ in the supersymmetric case by 
              using Eq.(\ref{massgap}) and making the choice 
              $g_0$ = 1.228 (and also N=3).  This would
              give for the mass of the lightest supersymmetric multiplet, 
                
                 \begin{equation}
          M^{\prime}/2 \approx 0.49~ {\rm GeV}\ .
                    \label{lightestsusygb}
          \end{equation}
          
                Clearly, the supersymmetric and non supersymmetric Yang-Mills model 
                results seem to be only qualitatively similar, the lightest supersymmetric 
                particle having a mass about 1/3 that of the
                non-supersymmetric model glueball (assuming the same coupling constant).

\begin{figure*}[t]
\begin{center}
\subfigure[~  Running of the scale as function of $g$ for Super Yang-Mills. ]
{\includegraphics[width=7.5cm]{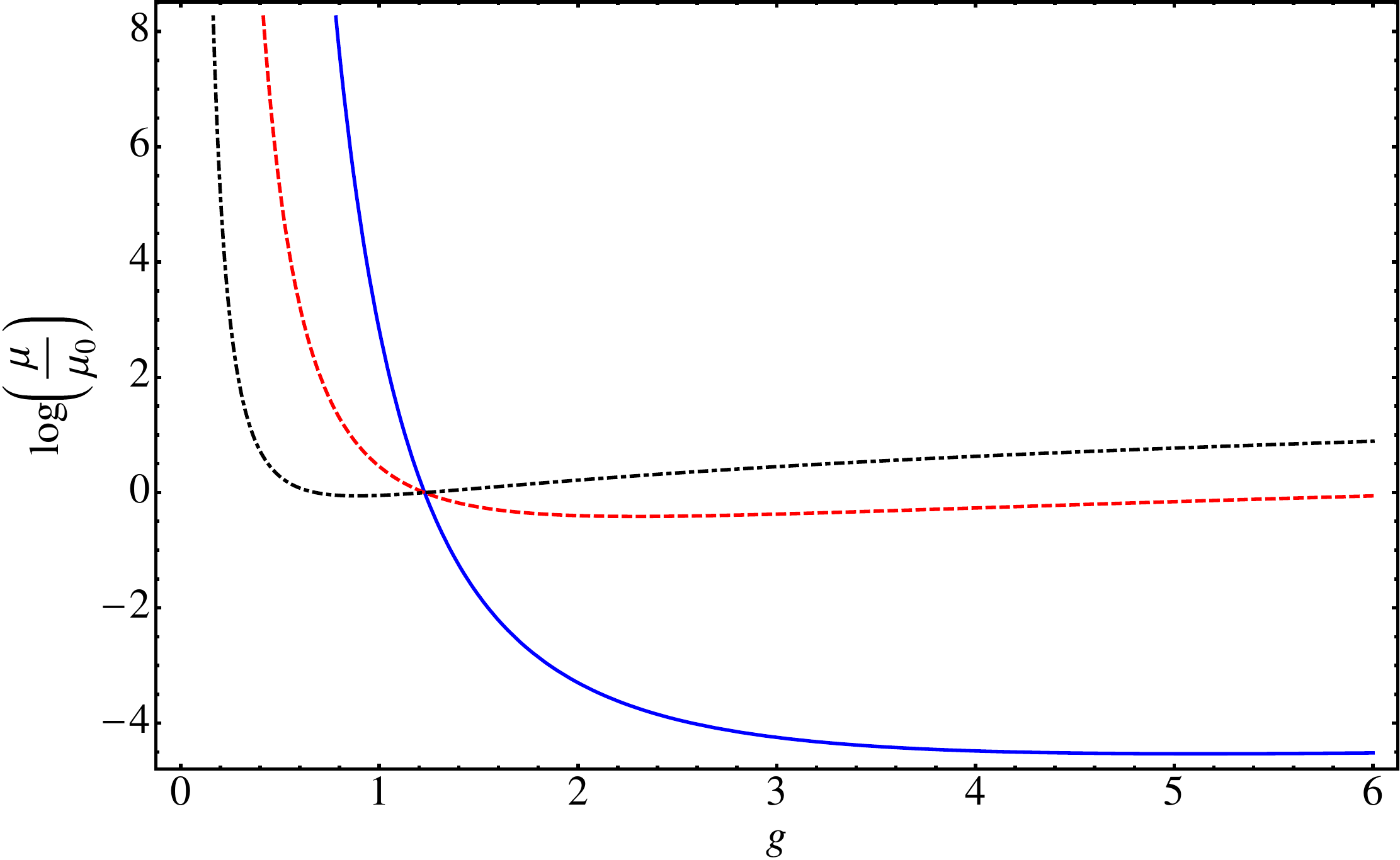}} 
\hskip 2cm
\subfigure[~Mass gap dependence on  $N$ for Super Yang-Mills ]{\includegraphics[width=7.5cm] {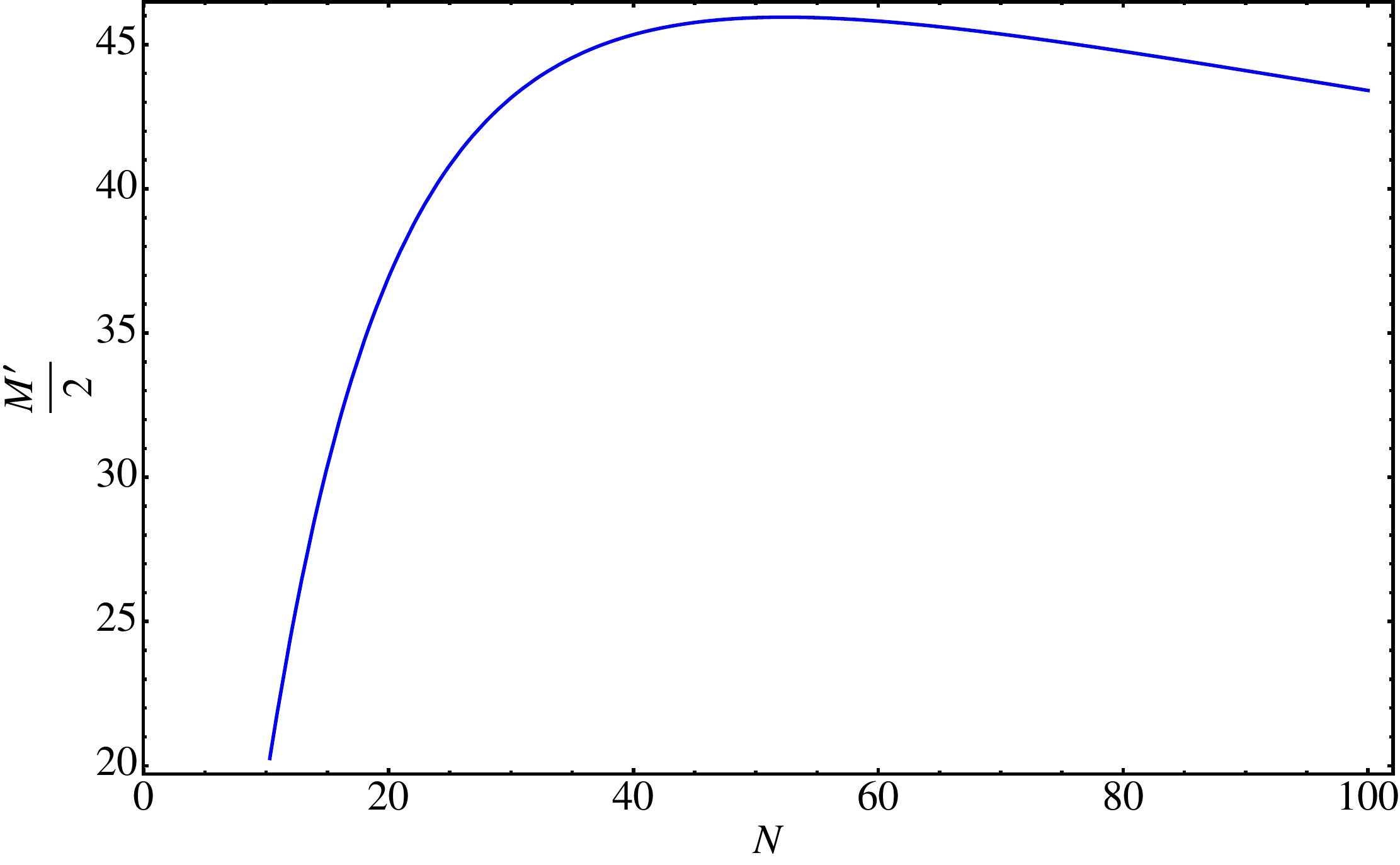}} \\
\caption{In the panel (a) we plot the solution of the running of the scale as function of the coupling constant $g$ for the Super Yang-Mills theory for $N=3$, solid blue line; $N=15$, red-dashed line and $N=100$, black-dot-dashed line.  We have kept fixed $g_0 = 1.228$ at the scale $\mu_0 = 91.19$~GeV. In the panel (b) we plot the mass gap $M^{\prime} /2$  as function of the number of colors $N$. } 
\label{FIGNdep}
\end{center}
\end{figure*}

               Of course, another interesting aspect to explore for gauge theories is their
               behavior as the number of colors, N gets large. Taking the supersymmetric
                gauge theory as an example and fixing, for the sake of definiteness, the gauge
                coupling constant as 1.228 at $\mu_0$ = 91.19 GeV, we find the running of 
                $g(\mu)$ by substituting Eq.(\ref{susyparam}) into Eq.(\ref{gruns}).
                Figure~\ref{FIGNdep}~(a) show $\log(\mu)$ plotted against $g(\mu)$
                for respectively N=3 (blue solid line), N=15 (red dashed line) and N=100 (black dot-dashed line). It is seen that the asymptotically free
                (left) region in $g$ shrinks as N increases. Furthermore, the mass gap 
                (corresponding
                to the ordinate at the minimum point) is shown in Fig.~~\ref{FIGNdep}~(b). It shows that the mass gap has a maximum as function of the number of colors.  
                
                 When considering the large N behavior of gauge theories it is often 
                 desired to make an extrapolation where large N is taken in such a 
                 way that the quantity,

                 \begin{equation}
                 g_t^2 \equiv g^2 N
                 \label{thcc}
                 \end{equation}
                  
                 is held fixed. With such an extrapolation there will be no
                 dependence (in either the supersymmetric Yang-Mills or 
                 the pure Yang-Mills with RS type beta function cases) on N.
                 This may be immediately seen by writing, from Eq.(\ref{RS}),
                 $a\equiv {\tilde a}N$ and 
                 $b\equiv {\tilde b}N$ so that
                 ${\tilde a}$ and ${\tilde b}$ are independent of N.
                 Then, as an example, Eq.(\ref{massgap}) for the mass gap becomes
                 
                 \begin{equation}
                 \log\left(\frac{M}{\mu_0}\right)= \frac{{\tilde b}}{{\tilde a}}
                 \left(\frac{1}{2} - \log({\tilde b}^{1/2}g_0 N^{1/2})\right) -
                 \frac{1}{2{\tilde a}g_0^2 N} \ ,
                 \label{bigNmassgap}
                 \end{equation}
                 
                 which only involves N via the fixed 
                 combination $g_0^2 N$.
                  
\section{Summary and discussion}

     First, an exploration of the scale dependence of physical quantities (with engineering
     dimension, $D$)  was made using dimensional analysis and a specific form, conjectured 
     by Ryttov and Sannino, of the beta function for the pure Yang-Mills theory. It was noted that
     even though the beta function had a pole (inherited from the known form for supersymmetric 
     gauge theories which stimulated the conjecture) physical quantities remained smooth as the
     pole value of the coupling constant, $g$ was crossed. 
     
     For a more detailed understanding of how the coupling constant runs, the integration
      of the beta function was next carried out for the complete set of SU(3) Yang-Mills theories, i.e. 
      those corresponding to any choice of coupling constant, $g_0$ at a convenient reference 
      scale, $\mu_0= m_Z$. Then $g(\mu)$ was also seen to have a smooth behavior at the pole value of g.
      Most interesting is that (due to the existence of the pole) the curve of $g(\mu)$
      predicts a numerical value for the mass gap of the Yang-Mills theory, i.e. 
      the mass of the lightest glueball. The predicted value, 1.67 GeV,
      seems rather close to the one obtained from a lattice treatment of QCD in the valence or
      quenched approximation.    

      It is noted that if the reference value of the coupling constant, $g_0$ decreases below about 1 
      (the experimental value is 1.228) the mass gap drops very quickly.
      
      A similar treatment was carried out for the supersymmetric SU(3) Yang-Mills
       theory and the mass gap was calculated to be about 0.49 GeV.
      In this case, the form of the beta function is quite similar but is based on a known
       rather than a conjectured beta function.
       
       We plan to next investigate the extent to which this work can be carried out for 
       non-supersymmetric 
      gauge theories containing fermions. Evidently there are many interesting questions which remain.

\subsection*{Acknowledgments}

We would like to thank Claudio Pica for useful discussions and comments. The work of JS was supported in part by
the US DOE under Contract No. DE-FGG-02-85ER 40231; he would also like to 
thank the CP${^3}$-Origins group at the University of Southern Denmark, where this 
work was started, for their warm hospitality and partial support.

\end{document}